\pdfoutput=1
\documentclass[10pt,journal]{IEEEtran}

\usepackage[breaklinks,colorlinks,bookmarksopen,bookmarksnumbered]{hyperref}

\usepackage{moreverb}
\usepackage{verbatim}

\usepackage{acronym}
\usepackage{enumitem}
\usepackage{color}

\usepackage{parskip}

\usepackage{subfig}
\usepackage{latexsym}
\usepackage{tabularx}
\usepackage{longtable}
\usepackage{cite} 
\usepackage[sharp]{easylist}

\usepackage{siunitx}
\usepackage[usenames,dvipsnames]{xcolor}

\usepackage{amsmath}
\usepackage{amsfonts}
\usepackage{amssymb}
\usepackage{amsthm}

\usepackage{algorithmic}
\usepackage{algorithm}
\usepackage{cases}

\usepackage[colorinlistoftodos,disable]{todonotes}

\usepackage{styles/mathdefs}

\renewcommand{\eqref}[1]{(\ref{eq:#1})}
\newcommand{\secref}[1]{\S\ref{sec:#1}}

\newcommand{\figref}[1]{Fig.~\ref{fig:#1}}
\newcommand{\tabref}[1]{Table~\ref{tab:#1}}

\newcommand{\current}{T_{\mbox{current}}}

\DeclareMathOperator*{\argmax}{\arg\!\max}

\newcommand{\textm}{{\scshape\textsf{text}}}
\newcommand{\webm}{{\scshape\textsf{web}}}
\newcommand{\appm}{{\scshape\textsf{app}}}
\newcommand{\locationm}{{\scshape\textsf{location}}}

\begin{document}

\title{Active Authentication on Mobile Devices via Stylometry, Application Usage, Web Browsing, and GPS Location}


\author{
  \IEEEauthorblockN{Lex Fridman\IEEEauthorrefmark{1}, Steven Weber\IEEEauthorrefmark{2},
    Rachel Greenstadt\IEEEauthorrefmark{2}, Moshe Kam\IEEEauthorrefmark{3}}
  \\\IEEEauthorblockA{\IEEEauthorrefmark{1}Massachusetts Institute of Technology
    \\fridman@mit.edu}
  \\\IEEEauthorblockA{\IEEEauthorrefmark{2}Drexel University
    \\\{sweber@coe, greenie@cs\}.drexel.edu}
  \\\IEEEauthorblockA{\IEEEauthorrefmark{3}New Jersey Institute of Technology
    \\moshe.kam@njit.edu}
}

\maketitle

\begin{abstract}%
  Active authentication is the problem of continuously verifying the identity of a person based on behavioral aspects of
  their interaction with a computing device. In this study, we collect and analyze behavioral biometrics data from 200
  subjects, each using their personal Android mobile device for a period of at least 30 days. This dataset is novel
  in the context of active authentication due to its size, duration, number of modalities, and absence of restrictions
  on tracked activity. The geographical colocation of the subjects in the study is representative of a large
  closed-world environment such as an organization where the unauthorized user of a device is likely to be an insider
  threat: coming from within the organization. We consider four biometric modalities: (1) text entered via soft
  keyboard, (2) applications used, (3) websites visited, and (4) physical location of the device as determined from GPS
  (when outdoors) or WiFi (when indoors). We implement and test a classifier for each modality and organize the
  classifiers as a parallel binary decision fusion architecture. We are able to characterize the performance of the
  system with respect to intruder detection time and to quantify the contribution of each modality to the overall
  performance.
\end{abstract}














\section{Introduction}

According to a 2013 Pew Internet Project study of 2076 people \cite{duggan2013cell}, 91\% of American adults own
a cellphone. Increasingly, people are using their phones to access and store sensitive data. The
same study found that 81\% of cellphone owners use their mobile device for texting, 52\% use it for email, 49\% use it
for maps (enabling location services), and 29\% use it for online banking. And yet, securing the data is often not taken
seriously because of an inaccurate estimation of risk as discussed in \cite{egelman2014you}. In particular, several
studies have shown that a large percentage of smartphone owners do not lock their phone: 57\% in \cite{harbach2014sa},
33\% in \cite{van2013modifying}, 39\% in \cite{egelman2014you}, and 48\% in this study.

Active authentication is an approach of monitoring the behavioral biometric characteristics of a user's interaction with
the device for the purpose of securing the phone when the point-of-entry locking mechanism fails or is absent.  In
recent years, continuous authentication has been explored extensively on desktop computers, based either on a single
biometric modality like mouse movement \cite{shen2012effectiveness} or a fusion of multiple modalities like keyboard
dynamics, mouse movement, web browsing, and stylometry \cite{fridman2013decision}. Unlike physical biometric devices
like fingerprint scanners or iris scanners, these systems rely on computer interface hardware like the keyboard and
mouse that are already commonly available with most computers.

In this paper, we consider the problem of active authentication on mobile devices, where the variety of available sensor
data is much greater than on the desktop, but so is the variety of behavioral profiles, device form factors, and
environments in which the device is used. Active authentication is the approach of verifying a user's identity
continuously based on various sensors commonly available on the device. We study four representative modalities of
stylometry (text analysis), application usage patterns, web browsing behavior, and physical location of the
device. These modalities were chosen, in part, due to their relatively low power consumption. In the remainder of the
paper these four modalities will be referred to as \textm{}, \appm{}, \webm{}, and \locationm{}, respectively. We
consider the trade-off between intruder detection time and detection error as measured by false accept rate (FAR) and
false reject rate (FRR). The analysis is performed on a dataset collected by the authors of 200 subjects using their
personal Android mobile device for a period of at least 30 days. To the best of our knowledge, this dataset is the first
of its kind studied in active authentication literature, due to its large size \cite{derawi2010unobtrusive}, the
duration of tracked activity \cite{li2014active}, and the absence of restrictions on usage patterns and on the form
factor of the mobile device. The geographical colocation of the participants, in particular, makes the dataset a good
representation of an environment such as a closed-world organization where the unauthorized user of a particular device
will most likely come from inside the organization.

We propose to use decision fusion in order to asynchronously integrate the four modalities and make serial
authentication decisions. While we consider here a specific set of binary classifiers, the strength of our
decision-level approach is that additional classifiers can be added without having to change the basic fusion
rule. Moreover, it is easy to evaluate the marginal improvement of any added classifier to the overall performance of
the system. We evaluate the multimodal continuous authentication system by characterizing the error rates of local
classifier decisions, fused global decisions, and the contribution of each local classifier to the fused decision. The
novel aspects of our work include the scope of the dataset, the particular portfolio of behavioral biometrics in the
context of mobile devices, and the extent of temporal performance analysis.

The remainder of the paper is structured as follows. In \secref{related-work}, we discuss the related work on multimodal
biometric systems, active authentication on mobile devices, and each of the four behavioral biometrics considered in
this paper. In \secref{dataset}, we discuss the 200 subject dataset that we collected and analyzed. In
\secref{algorithms}, we discuss four biometric modalities, their associated classifiers, and the decision fusion
architecture. In \secref{results}, we present the performance of each individual classifier, the performance of the
fusion system, and the contribution of each individual classifier to the fused decisions.

\section{Related Work}\label{sec:related-work}

\subsection{Multimodal Biometric Systems}

The window of time based on which an active authentication system is tasked with making a binary decision is relatively 
short and thus contains a highly variable set of biometric information. Depending on the task the user is engaged in,
some of the biometric classifiers may provide more data than others. For example, as the user chats with a friend via
SMS, the text-based classifiers will be actively flooded with data, while the web browsing based classifiers may only
get a few infrequent events. This motivates the recent work on multimodal authentication systems where the decisions of
multiple classifiers are fused together \cite{sim2007continuous}. In this way, the verification process is more robust
to the dynamic nature of human-computer interaction. The current approaches to the fusion of classifiers
center around max, min, median, or majority vote combinations \cite{kittler1998combining}. When neural networks are used
as classifiers, an ensemble of classifiers is constructed and fused based on different initialization of the neural
network \cite{chen2013optimal}.

Several active authentication studies have utilized multimodal biometric systems but have all, to the best of our
knowledge: (1) considered a smaller pool of subjects, (2) have not characterized the temporal performance of intruder
detection, and (3) have shown overall significantly worse performance than that achieved in our study. 

Our approach in this paper is to apply the Chair-Varshney optimal fusion rule \cite{chair-varshney-1986-optimal} for the
combination of available multimodal decisions. The strength of the decision-level fusion approach is that an arbitrary
number of classifiers can be added without re-training the classifiers already in the system. This modular design allows
for multiple groups to contribute drastically different classification schemes, each lowering the error rate of the
global decision.

\subsection{Mobile Active Authentication}

With the rise of smartphone usage, active authentication on mobile devices has begun to be studied in the last few
years. The large number of available sensors makes for a rich feature space to explore. Ultimately, the question is the
one that we ask in this paper: what modality contributes the most to a decision fusion system toward the goal of fast,
accurate verification of identity? Most of the studies focus on a single modality. For example, gait pattern was
considered in \cite{derawi2010unobtrusive} achieving an EER of 0.201 (20.1\%) for 51 subjects during two short sessions,
where each subject was tasked with walking down a hallway. Some studies have incorporated multiple modalities. For
example, keystroke dynamics, stylometry, and behavioral profiling were considered in \cite{saevanee2014text} achieving
an EER of 0.033 (3.3\%) from 30 simulated users. The data for these users was pieced together from different
datasets. To the best of our knowledge, the dataset that we collected and analyzed is unique in all its key aspects: its
size (200 subjects), its duration (30+ days), and the size of the portfolio of modalities that were all tracked
concurrently with a synchronized timestamp.

\subsection{Stylometry, Web Browsing, Application Usage, Location}

Stylometry is the study of linguistic style. It has been extensively applied to the problems of authorship attribution,
identification, and verification. See \cite{brocardo2013authorship} for a thorough summary of stylometric studies in
each of these three problem domains along with their study parameters and the resulting accuracy. These studies
traditionally use large sets of features (see Table II in \cite{abbasi2008writeprints}) in combination with support
vector machines (SVMs) that have proven to be effective in high dimensional feature space \cite{mcdonald:12}, even in
cases when the number of features exceeds the number of samples. Nevertheless, with these approaches, often more than
500 words are required in order to achieve adequately low error rates \cite{fridman2014multimodal}. This makes them
impractical for the application of real-time active authentication on mobile devices where text data comes in short
bursts. While the other three modalities are not well investigated in the context of active authentication, this is not
true for stylometry. Therefore, for this modality, we don't reinvent the wheel, and implement the n-gram analysis
approach presented in \cite{brocardo2013authorship} that has been shown to work sufficiently well on short blocks of
texts.

Web browsing, application usage, and location have not been studied extensively in the context of active
authentication. The following is a discussion of the few studies that we are aware of. Web browsing behavior has been
studied for the purpose of understanding user behavior, habits, and interests \cite{yampolskiy2008behavioral}. Web
browsing as a source for behavioral biometric data was considered in \cite{abramson2013user} to achieve average
identification FAR/FRR of 0.24 (24\%) on a dataset of 14 desktop computer users.  Application usage was considered in
\cite{li2014active}, where cellphone data (from 2004) from the MIT Reality Mining project \cite{eagle2009inferring} was
used to achieve 0.1 (10\%) EER based on a portfolio of metrics including application usage, call patterns, and
location. Application usage and movements patterns have been studied as part of behavioral profiling in cellular
networks \cite{sun2004mobility, hall2005anomaly, li2014active}. However, these approaches use position data of lower
resolution in time and space than that provided by GPS on smartphones. To the best of our knowledge, GPS traces have not
been utilized in literature for continuous authentication.

\section{Dataset}\label{sec:dataset}

The dataset used in this work contains behavioral biometrics data for 200 subjects. The collection of the data was
carried out by the authors over a period of 5 months. The requirements of the study were that each subject was a student
or employee of Drexel University and was an owner and an active user of an Android smartphone or tablet. The number of
subjects with each major Android version and associated API level are listed in \tabref{android-versions}. Nexus 5 was
the most popular device with 10 subjects using it. Samsung Galaxy S5 was the second most popular device with 6 subjects
using it.

\begin{table}[h]
  \centering
  \begin{tabular}{| l | l | l |}
    \hline
    \textbf{Android Version} & \textbf{API Level} & \textbf{Subjects} \\\hline
    4.4 & 19 & 143\\\hline
    4.1 & 16 & 16\\\hline
    4.3 & 18 & 15\\\hline
    4.2 & 17 & 9\\\hline
    4.0.4 & 15 & 5\\\hline
    2.3.6 & 10 & 4\\\hline
    4.0.3& 15 & 3\\\hline
    2.3.5& 10 & 3\\\hline
    2.2 & 8 & 2\\\hline
  \end{tabular}
  \caption{The Android version and API level of the 200 devices that were part of the study.}
  \label{tab:android-versions}
\end{table}

A tracking application was installed on each subject's device and operated for a period of at least 30 days until the
subject came in to approve the collected data and get the tracking application uninstalled from their device.  The
following data modalities were tracked with 1-second resolution:

\begin{itemize}
\item Text typed via soft keyboard.
\item Apps visited.
\item Websites visited.
\item Location (based on GPS or WiFi).
\end{itemize}

The key characteristics of this dataset are its large size (200 users), the duration of tracked activity (30+ days), and
the geographical colocation of its participants in the Philadelphia area. Moreover, we did not place any restrictions on
usage patterns, on the type of Android device, and on the Android OS version (see \tabref{android-versions}).

There were several challenges encountered in the collection of the data. The biggest problem was battery drain. Due to
the long duration of the study, we could not enable modalities whose tracking proved to be significantly draining of battery
power. These modalities include front-facing video for eye tracking and face recognition, gyroscope, accelerometer, and
touch gestures. Moreover, we had to reduce GPS sampling frequency to once per minute on most of the devices.

\begin{table}[h]
  \centering
  \begin{tabular}{| l | l | }
    \hline
    \textbf{Event} & \textbf{Frequency} \\\hline
    Text & 23,254,478\\\hline
    App & 927,433\\\hline
    Web & 210,322\\\hline
    Location & 143,875\\\hline
  \end{tabular}
  \caption{The number of events in the dataset associated with each of the four modalities considered in this
    paper. A \textm{} event refers to a single character entered on the soft keyboard. An \appm{} events refers to a new
    app receiving focus. A \webm{} event refers to a new url entered in the url box. A \locationm{} event refers to a
    new sample of the device location either from GPS or WiFi.}
  \label{tab:event-count}
\end{table}

\tabref{event-count} shows statistics on each of the four investigated modalities in the corpus. The table contains data
aggregated over all 200 users. The ``frequency'' here is a count of the number of instances of an action associated with
that modality. As stated previously, the four modalities will be referred to as \textm{}, \appm{}, \webm{}, and
``location.''  For \textm{}, the action is a single keystroke on the soft keyboard. For \appm{}, the action is opening
or bringing focus to a new app. For \webm{}, the action is visiting a new website. For \locationm{}, no explicitly
action is taken by the user. Rather, location is sampled regularly at intervals of 1 minute when GPS is enabled. As
\tabref{event-count} suggests, \textm{} events fire 1-2 orders of magnitude more frequently than the other
three.

\begin{figure}[h]
  \centering
  \includegraphics[width=\columnwidth]{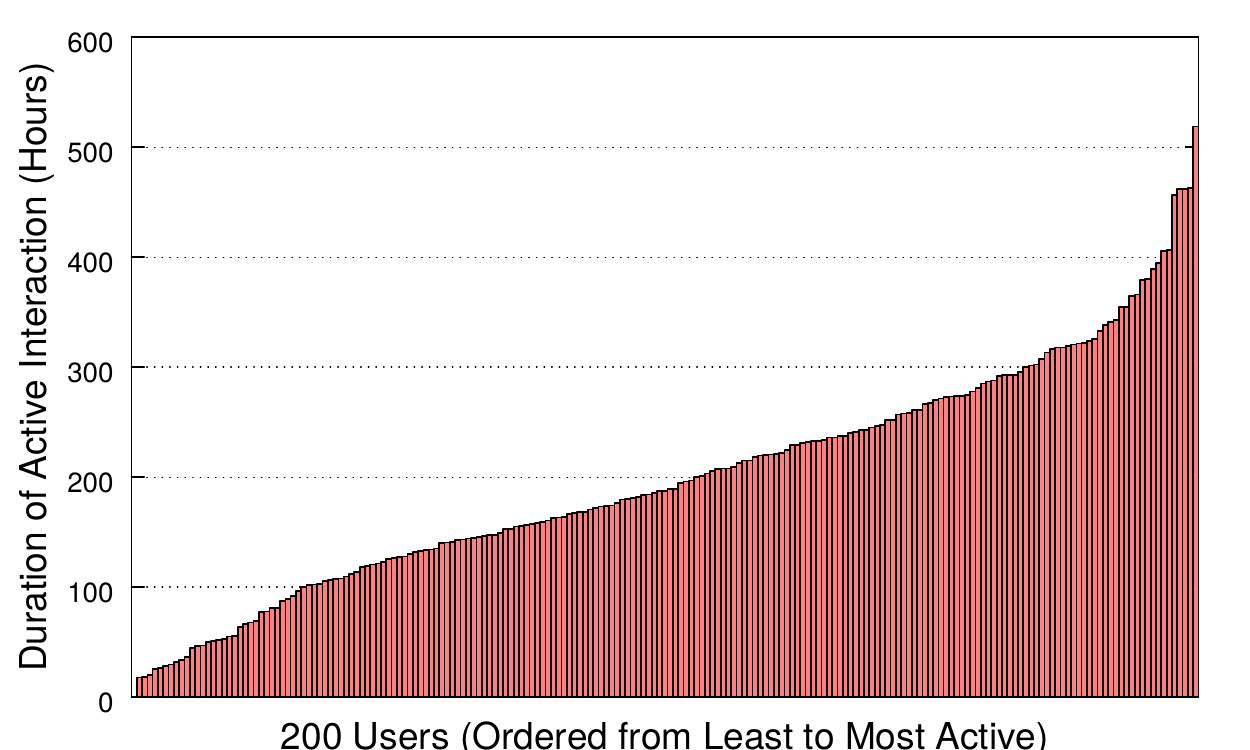}
  \caption{The duration of time (in hours) that each of the 200 users actively interacted with their device..}
  \label{fig:active-duration}
\end{figure}

The data for each user is processed to remove idle periods when the device is not active. The threshold for what is
considered an idle period is 5 minutes. For example, if the time between event A and event B is 20 minutes, with no
other events in between, this 20 minutes is compressed down to 5 minutes. The date and time of the event are not changed
but the timestamp used in dividing the dataset for training and testing (see \secref{training-char-testing}) is updated
to reflect the new time between event A and event B. This compression of idle times is performed in order to regularize
periods of activity for cross validation that utilizes time-based windows as described in
\secref{training-char-testing}. The resulting compressed timestamps are referred to as ``active
interaction''. \figref{active-duration} shows the duration (in hours) of active interaction for each of the 200 users
ordered from least to most active.

\begin{table*}[!htbp]
  \centering
  \subfloat[]{\label{tab:keys-per-app}
    \begin{tabular}{| l | l | }
      \hline
      \textbf{App Name} & \textbf{Keys Per App} \\\hline
      com.android.sms & 5,617,297\\\hline
      com.android.mms & 5,552,079\\\hline
      com.whatsapp & 4,055,622\\\hline
      com.facebook.orca & 1,252,456\\\hline
      com.google.android.talk & 1,147,295\\\hline
      com.infraware.polarisviewer4 & 990,319\\\hline
      com.android.chrome & 417,165\\\hline
      com.facebook.katana & 405,267\\\hline
      com.snapchat.android & 377,840\\\hline
      com.google.android.gm & 271,570\\\hline
      com.htc.sense.mms & 238,300\\\hline
      com.tencent.mm & 221,461\\\hline
      com.motorola.messaging & 203,649\\\hline
      com.android.calculator2 & 167,435\\\hline
      com.verizon.messaging.vzmsgs & 137,339\\\hline
      com.groupme.android & 134,896\\\hline
      com.handcent.nextsms & 123,065\\\hline
      com.jb.gosms & 118,316\\\hline
      com.sonyericsson.conversations & 114,219\\\hline
      com.twitter.android & 92,605\\\hline
    \end{tabular}
  }
  \hspace{0.3in}
  \subfloat[]{\label{tab:top-app}
    \begin{tabular}{| l | l | }
      \hline
      \textbf{App Name} & \textbf{Visits} \\\hline
      TouchWiz home & 101,151\\\hline
      WhatsApp & 64,038\\\hline
      Messaging & 60,015\\\hline
      Launcher & 39,113\\\hline
      Facebook & 38,591\\\hline
      Google Search & 32,947\\\hline
      Chrome & 32,032\\\hline
      Snapchat & 23,481\\\hline
      System UI & 22,772\\\hline
      Phone & 19,396\\\hline
      Gmail & 19,329\\\hline
      Messages & 19,154\\\hline
      Contacts & 18,668\\\hline
      Hangouts & 17,209\\\hline
      Home & 16,775\\\hline
      HTC Sense & 16,325\\\hline
      YouTube & 14,552\\\hline
      Xperia Home & 13,639\\\hline
      Instagram & 13,146\\\hline
      Settings & 12,675\\\hline
    \end{tabular}
  }
  \hspace{0.3in}
  \subfloat[]{\label{tab:top-web}
    \begin{tabular}{| l | l | }
      \hline
      \textbf{Website Domain} & \textbf{Visits} \\\hline
      www.google.com & 19,004\\\hline
      m.facebook.com & 9,300\\\hline
      www.reddit.com & 4,348\\\hline
      forums.huaren.us & 3,093\\\hline
      learn.dcollege.net & 2,133\\\hline
      en.m.wikipedia.org & 1,825\\\hline
      mail.drexel.edu & 1,520\\\hline
      one.drexel.edu & 1,472\\\hline
      login.drexel.edu & 1,462\\\hline
      likes.com & 1,361\\\hline
      mail.google.com & 1,292\\\hline
      i.imgur.com & 1,132\\\hline
      www.amazon.com & 1,079\\\hline
      netcontrol.irt.drexel.edu & 1,049\\\hline
      www.facebook.com & 903\\\hline
      banner.drexel.edu & 902\\\hline
      m.hupu.com & 824\\\hline
      t.co & 801\\\hline
      duapp2.drexel.edu & 786\\\hline
      m.ign.com & 725\\\hline
    \end{tabular}
  }
  \caption{Top 20 apps ordered by text entry and visit frequency and top 20 websites ordered by visit frequency. These
    tables are provided to give insight into the structure and content of the dataset.}
  \label{tab:top-20}
\end{table*}

\tabref{top-20} shows three top-20 lists: (1) the top-20 apps based on the amount of text that was typed inside each
app, (2) the top-20 apps based on the number of times they received focused, and (3) the top-20 website domains based on
the number of times a website associated with that domain was visited. These are aggregate measures across the dataset
intended to provide an intuition about its structure and content, but the top-20 list is the same as that used for the
the classifier model based on the \webm{} and \appm{} features in \secref{algorithms}.

\begin{figure}[h]
  \centering
  \includegraphics[width=\columnwidth]{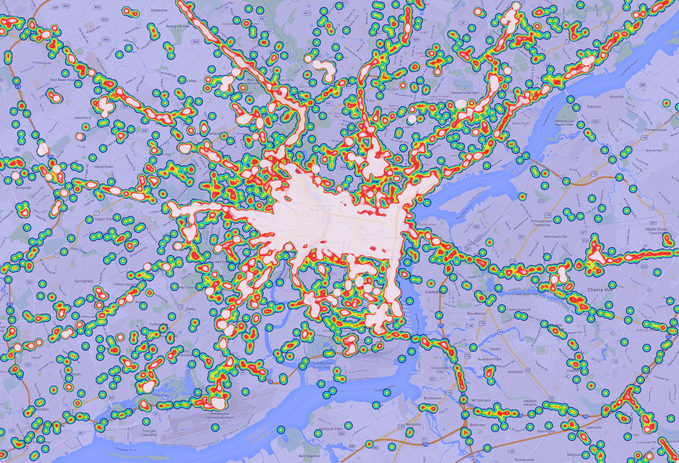}
  \caption{An aggregate heatmap showing a selection from the dataset of GPS locations in the Philadelphia area.}
  \label{fig:location-heatmap}
\end{figure}

\figref{location-heatmap} shows a heat map visualization of a selection from the dataset of GPS locations in the
Philadelphia area. The subjects in the study resided in Philadelphia but traveled all over United States and the
world. There are two key characteristics of the GPS location data. First, it is relatively unique to each individual
even for people living in the same area of a city. Second, outside of occasional travel, it does not vary significantly
from day to day. Human beings are creatures of habit, and in as much as location is a measure of habit, this idea is
confirmed by the location data of the majority of the subjects in the study.

\section{Classification and Decision Fusion}\label{sec:algorithms}

\subsection{Features and Classifiers}

The four distinct biometric modalities considered in our analysis are (1) text entered via soft keyboard, (2)
applications used, (3) websites visited, and (4) physical location of the device as determined from GPS (when outdoors)
or WiFi (when indoors). We refer to these four modalities as \textm{}, \appm{}, \webm{}, and \locationm{},
respectively. In this section we discuss the features that were extracted from the raw data of each modality, and the
classifiers that were used to map these features into binary decision space.

A binary classifier is constructed for each of the 200 users and 4 modalities. In total, there are 800 classifiers, each
producing either a probability that a user is valid $P(H_1)$ (or a binary decision of 0 (invalid) or 1 (valid). The
first class ($H_1$) for each classifier is trained on the valid user's data and the second class ($H_0$) is trained on
the other 199 users' data. The training process is described in more detail in \secref{training-char-testing}. For
\appm{}, \webm{}, and \locationm{}, the classifier takes a single instance of the event and produces a probability. For multiple
events of the same modality, the set of probabilities is fused across time using maximum likelihood:

\begin{equation}\label{eq:max-likelihood}
H^* = \argmax_{i\in\{0,1\}} \prod_{x_t\in \Omega}P(x_t|H_i), 
\end{equation}

\noindent where $\Omega = \{x_t | \current - T(x_t) \leq \omega\}$, $\omega$ is a fixed window size in seconds,
$T(x_t)$ is the timestamp of event $x_t$, and $\current$ is the current timestamp. The process of fusing classifier
scores across time is illustrated in \figref{fusion-across-time-and-classifiers}.

\subsubsection{Text}

As \tabref{keys-per-app} indicates, the apps into which text was entered on mobile devices varied, but the activity in
majority of the cases was communication via SMS, MMS, WhatsApp, Facebook, Google Hangouts, and other chat
apps. Therefore, \textm{} events fired in short bursts. The tracking application captured the keys that were
touched on the keyboard and not the autocorrected result. Therefore, the majority of the typed messages had a lot of
misspellings and words that were erased in the final submitted message. In the case of SMS, we also were able to record
the submitted result. For example, an SMS text that was submitted as ``\texttt{Sorry couldn't call back.}'' had
associated with it the following recorded keystrokes: ``\texttt{Sprry coyld cpuldn't vsll back.}'' Classification based
on the actual typed keys in principle is a better representation of the person's linguistic style. It captures unique typing
idiosyncrasies that autocorrect can conceal. As discussed in \secref{related-work}, we implemented a one-feature n-gram
classifier from \cite{brocardo2013authorship} that has been shown to work well on short messages. It works by
analyzing the presence or absence of n-grams with respect to the training set.

\subsubsection{App and Web}

The \appm{} and \webm{} classifier models we construct are identical in their structure. For the \appm{} modality we use the
app name as the unique identifier and count the number of times a user visits each app in the training set. For the \webm{}
modality we use the domain of the URL as the unique identifier and count the number of times a user visits each domain
in the training set. Note that, for example, ``m.facebook.com'' is a considered a different domain than
``www.facebook.com'' because the subdomain is different. In this section we refer to the app name and the web domain as
an ``entity''. \tabref{top-app} and \tabref{top-web} show the top entities aggregated across all 200 users for \appm{}
and \webm{} respectively.

For each user, the classification model for the valid class is constructed by determining the top 20 entities visited by
that user in the training set. The quantity of visits is then normalized so that the 20 frequency values sum to 1. The
classification model for the invalid class is constructed by counting the number of visit by the other 199 users to
those same 20 domains, such that for each of those domains we now have a probability that a valid user visits it and an
invalid user visits it. The evaluation for each user given the two empirical distributions is performed by the maximum
likelihood product in \eqref{max-likelihood}. Entities that do not appear in the top 20 are considered outliers and are
ignored in this classifier.

\subsubsection{Location}

Location is specified as a pair of values: latitude and longitude. Classification is performed using support vector
machines (SVMs) \cite{abe2010support} with the radial basis function (RBF) as the kernel function. The SVM produces a
classification score for each pair of latitude and longitude. This score is calibrated to form a probability using Platt
scaling \cite{niculescu2005predicting} which requires an extra logistic regression on the SVM scores via an additional
cross-validation on the training data. All of the code in this paper is written by the authors except for the SVM
classifier. Since the authentication system is written in C++, we used the Shark 3.0 machine learning library for the
SVM implementation.

\subsection{Decision Fusion}\label{sec:decision-fusion}

\begin{figure*}[!htbp]
  \centering
  \includegraphics[width=\textwidth]{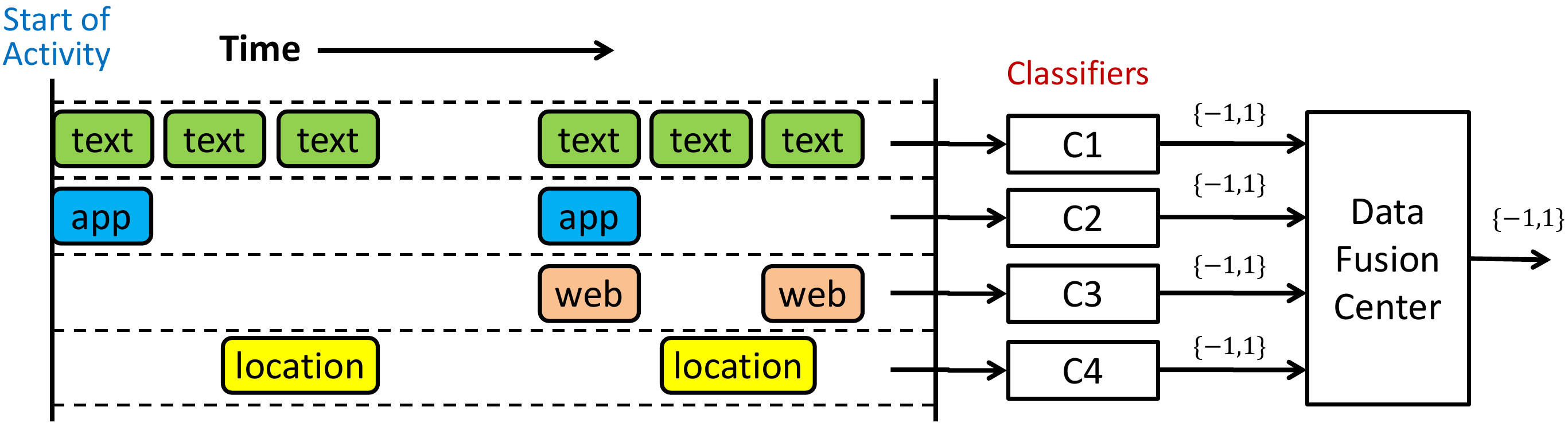}
  \caption{The fusion architecture across time and across classifiers. The \textm{}, \appm{}, \webm{}, and \locationm{}
    boxes indicate a firing of a single event associated with each of those modalities. Multiple classifier scores from
    the same modality are fused via \eqref{max-likelihood} to produce a single local binary decision. Local binary
    decisions from each of the four modalities are fused via \eqref{fusion-rule} to produce a single global binary
    decision.}
  \label{fig:fusion-across-time-and-classifiers}
\end{figure*}

Decision fusion with distributed sensors is described by Tenney and Sandell in \cite{Tenney1981} who studied a parallel
decision architecture. As described in \cite{Kam1991}, the system comprises of $n$ local detectors, each making a
decision about a binary hypothesis $(H_0,H_1)$, and a decision fusion center (DFC) that uses these local decisions
$\{u_1,u_2,...,u_n \}$ for a global decision about the hypothesis. The $i^{th}$ detector collects $K$ observations
before it makes its decision, $u_i$. The decision is $u_i =1$ if the detector decides in favor of $H_1$ and $u_i = -1$
if it decides in favor of $H_0$. The DFC collects the $n$ decisions of the local detectors and uses them in order to
decide in favor of $H_0 (u = -1)$ or in favor of $H_1 (u = 1)$.  Tenney and Sandell \cite{Tenney1981} and Reibman and
Nolte \cite{Reibman1987} studied the design of the local detectors and the DFC with respect to a Bayesian cost, assuming
the observations are independent conditioned on the hypothesis. The ensuing formulation derived the local and DFC
decision rules to be used by the system components for optimizing the system-wide cost. The resulting design requires
the use of likelihood ratio tests by the decision makers (local detectors and DFC) in the system. However the thresholds
used by these tests require the solution of a set of nonlinear coupled differential equations. In other words, the
design of the local decision makers and the DFC are co-dependent. In most scenarios the resulting complexity renders the
quest for an optimal design impractical.

Chair and Varshney in \cite{chair-varshney-1986-optimal} developed the optimal fusion rule when the local detectors are
fixed and local observations are statistically independent conditioned on the hypothesis. Data Fusion Center is optimal
given the performance characteristics of the local fixed decision makers. The result is a suboptimal (since local
detectors are fixed) but computationally efficient and scalable design. In this study we use the Chair-Varshney
formulation. The parallel distributed fusion scheme (see \figref{fusion-across-time-and-classifiers}) allows each
classifier to observe an event, minimize the local risk and make a local decision over the set of hypothesis, based on
only its own observations. Each classifier sends out a decision of the form:

\begin{equation}
  u_i =
  \begin{cases}
    1, & \mbox{if $H_1$ is decided}\\
    -1, & \mbox{if $H_0$ is decided}
  \end{cases}
\end{equation}

The fusion center combines these local decisions by minimizing the global Bayes' risk. The optimum decision rule
performs the following likelihood ratio test
\begin{eqnarray}\label{eq:Chair_Varshney_LRT}
\frac{P(u_1,...,u_n|H_1)}{P(u_1,...,u_n|H_0)} \underset{H_0}{\overset{H_1}{\gtrless}}  \frac{P_0}{P_1} = \tau
\end{eqnarray}
where the a priori probabilities of the binary hypotheses $H_1$ and $H_0$ are $P_1$ and $P_0$ respectively. In this case
the general fusion rule proposed in \cite{chair-varshney-1986-optimal} is

\begin{equation}\label{eq:fusion-rule}
  f(u_1, ..., u_n) =
  \begin{cases}
    1, & \mbox{if } a_0+\sum_{i=0}^n a_i u_i > 0\\
    -1, & \mbox{otherwise}
  \end{cases}
\end{equation}
with $P^M_i , P^F_i$ representing the \emph{False Rejection Rate} (FRR) and \emph{False Acceptance Rate} (FAR) of the $i^{th}$ classifier respectively. The optimum weights minimizing the global probability of error are given by

\begin{align}
  a_0 &= \log{\frac{P_1}{P_0}} \label{eq:const_weight}\\
  a_i &= \label{eq:nonconst_weight}
  \begin{cases}
    \log{\frac{1-P^M_i}{P^F_i}}, & \mbox{if } u_i=1\\
    \log{\frac{1-P^F_i}{P^M_i}}, & \mbox{if } u_i=-1
  \end{cases}
\end{align}

The threshold in \eqref{Chair_Varshney_LRT} requires knowledge of the a priori probabilities of the hypotheses. In
practice, these probabilities are not available, and the threshold $\tau$ is determined using different considerations
such as fixing the probability of false alarm or false rejection as is done in \secref{results-fusion}.

\section{Results}\label{sec:results}

\subsection{Training, Characterization, Testing}\label{sec:training-char-testing}

\begin{figure}[h]
  \centering
  \includegraphics[width=\columnwidth]{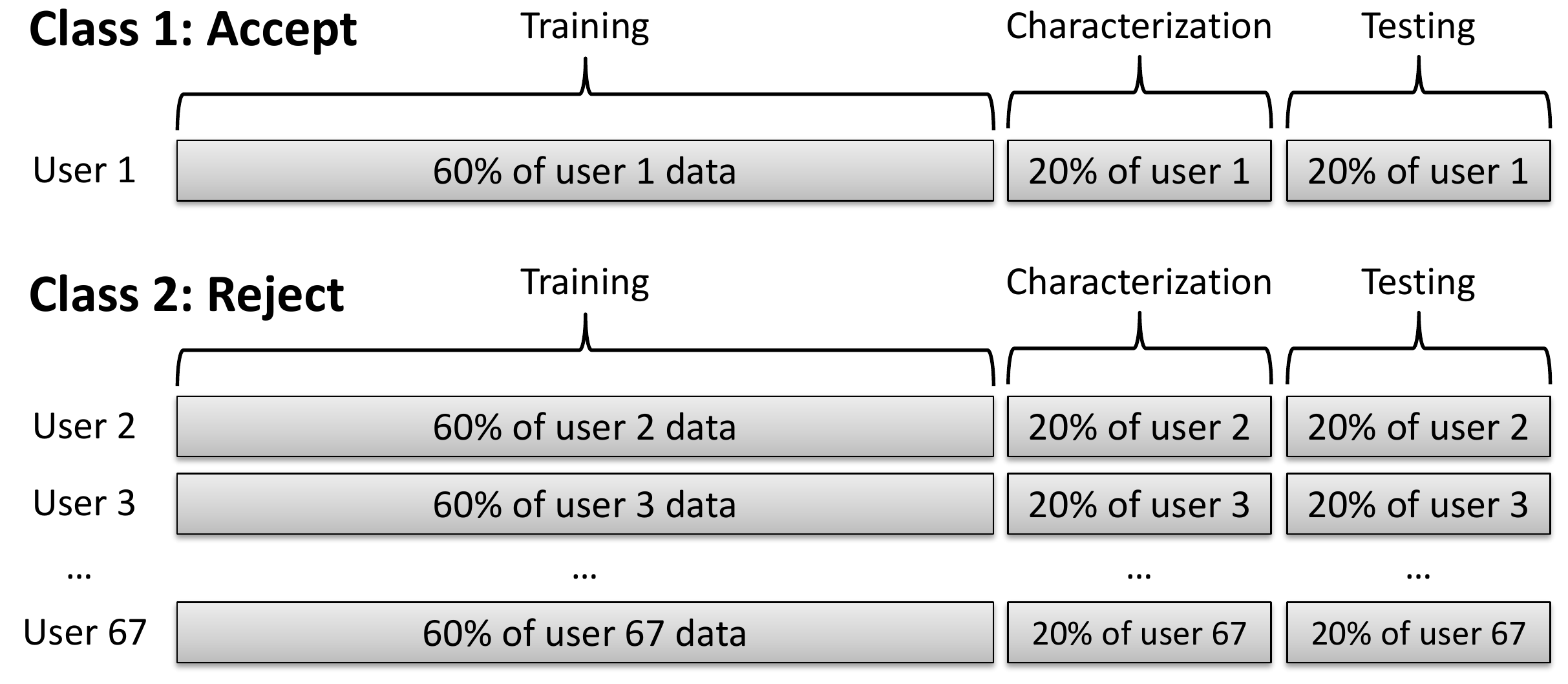}
  \caption{The three phases of processing the data to determine the individual performance of each classifiers and the
    performance of the fusion system that combines some subset of these classifiers.}
  \label{fig:training-char-testing-phases}
\end{figure}

The data of each of the 200 users' active interaction with the mobile device was divided into 5 equal-size folds (each
containing 20\% time span of the full set). We performed training of each classifier on the first three folds (60\%). We
then tested their performance on the fourth fold. This phase is referred to as ``characterization'', because its sole
purpose is to form estimates of FAR and FRR for use by the fusion algorithm. We then tested the performance of the
classifiers, individually and as part of the fusion system, on the fifth fold. This phase is referred to as ``testing''
since this is the part that is used for evaluation the performance of the individual classifiers and the fusion system. The
three phases of training, characterization, and testing as they relate to the data folds are shown in
\figref{training-char-testing-phases}.

\begin{itemize}
\item Training on folds 1, 2, 3.\\ Characterization on fold 4.\\ Testing on fold 5.
\item Training on folds 2, 3, 4.\\ Characterization on fold 5.\\ Testing on fold 1.
\item Training on folds 3, 4, 5.\\ Characterization on fold 1.\\ Testing on fold 2.
\item Training on folds 4, 5, 1.\\ Characterization on fold 2.\\ Testing on fold 3.
\item Training on folds 5, 1, 2.\\ Characterization on fold 3.\\ Testing on fold 4.
\end{itemize}

The common evaluation method used with each classifier for data fusion was measuring the averaged error rates across
five experiments; In each experiment, data of 3 folds was taken for training, 1 fold for characterization, and 1 for
testing. The FAR and FRR computed during characterization were taken as input for the fusion system as a measurement of
the expected performance of the classifiers. Therefore each experiment consisted of three phases: 1) train the
classifier(s) using the training set, 2) determine FAR and FRR based on the training set, and 3) classify the windows in
the test set.

\subsection{Performance: Individual Classifiers}\label{sec:results-individual}

\begin{figure}[h]
  \centering
  \includegraphics[width=\columnwidth]{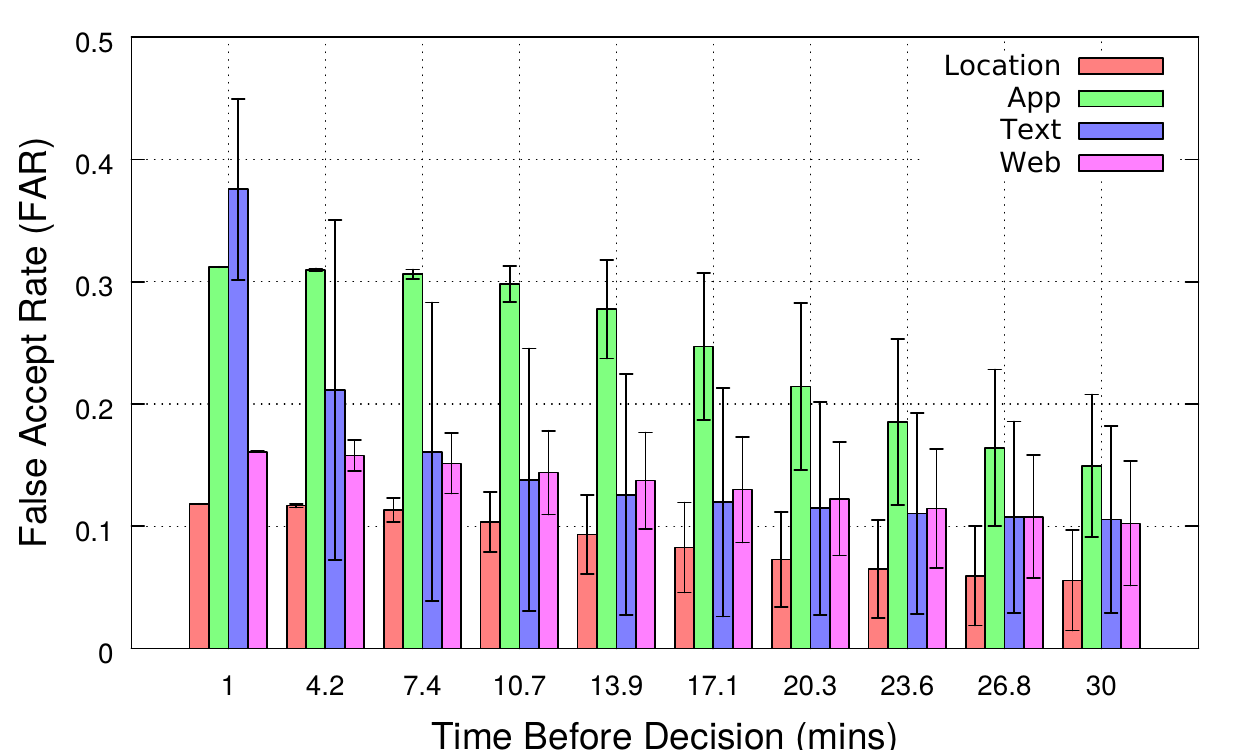}\\\vspace{0.2in}
  \includegraphics[width=\columnwidth]{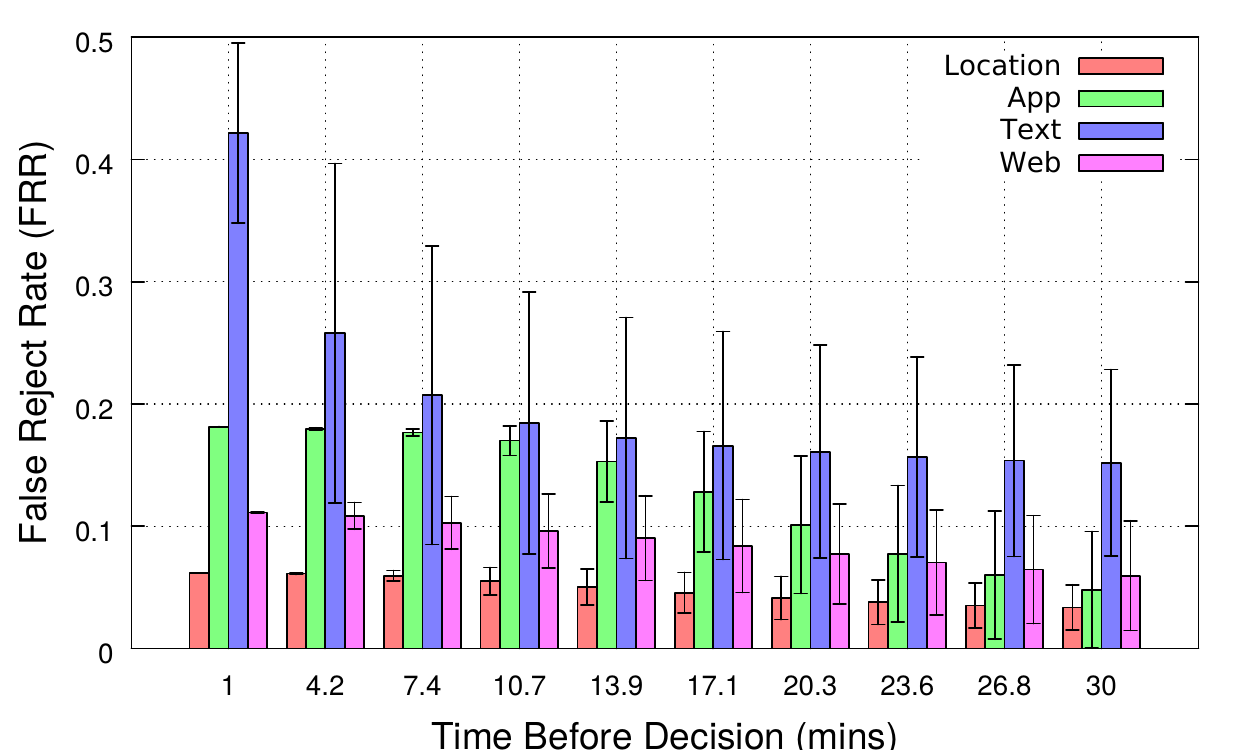}
  \caption{FAR and FRR performance of the individual classifiers associated with each of the four modalities. Each bar
    represent the average error rate for a given module and time window. Each of the 200 users has 2 classifiers for
    each modality, so each bar provides a value that was averaged over 200 individual error rates. The error bar
    indicate the standard deviation across these 200 values.}
  \label{fig:far-frr-individual}
\end{figure}

The conflicting objectives of an active authentication system are of response-time and performance. The less the system
waits before making an authentication decision, the higher the expected rate of error. As more behavioral biometric data
trickles in, the system can, on average, make a classification decision with greater certainty.

This pattern of decreased error rates with an increased decision window can be observed in \figref{far-frr-individual}
that shows (for 10 different time windows) the FAR and FRR of the 4 classifiers averaged over the 200 users with the
error bars indicating the standard deviation. The ``testing fold'' (see \secref{training-char-testing}) is used for
computing these error rates. The ``characterization fold'' does not affect these results, but is used only for FAR/FRR
estimation required by the decision fusion center in \secref{results-fusion}.

The ``time before decision'' is the time between the first event indicating activity and the first decision produced by the
fusion system. This metric can be thought of as ``decision window size''. Events older than the time range covered by
the time-window are disregarded in the classification. If no event associated with the modality under consideration
fires in a specific time window, no error is added to the average.

\begin{table}[h]
  \centering
  \begin{tabular}{| l | l | }
    \hline
    \textbf{Event} & \textbf{Firing Rate (per hour)}\\\hline
    Text & 557.8\\\hline
    App & 23.2\\\hline
    Web & 5.6\\\hline
    Location & 3.5\\\hline
  \end{tabular}
  \caption{The rates at which an event associated with each modality ``fires'' per hour. On average, GPS location is
    provided only 3.5 times an hour.}
  \label{tab:firing-rate}
\end{table}

There are two notable observations about the FAR/FRR plots in \figref{far-frr-individual}. First, the location modality
provides the lowest error rates even though on average across the dataset it fires only 3.5 times an hour as shown in
\tabref{firing-rate}. This means that classification on a single GPS coordinate is sufficient to correctly verify the
user with an FAR of under 0.1 and an FRR of under 0.05. Second, the text modality converges to an FAR of 0.16 and an FRR
of 0.11 after 30 minutes which is one of the worse performers of the four modalities, even though it fires 557.8 times
an hour on average. At the 30 minute mark, that firing rate equates to an average text block size of 279 characters. An
FAR/FRR of 0.16/0.11 with 279 characters blocks improves on the error rates achieved in \cite{brocardo2013authorship}
with 500 character blocks which in turn improved on the errors rates achieved in prior work for blocks of small text
(see \cite{brocardo2013authorship} for a full reference list on short-text stylometric analysis).

\subsection{Performance: Decision Fusion}\label{sec:results-fusion}

\begin{figure}[h]
  \centering
  \includegraphics[width=\columnwidth]{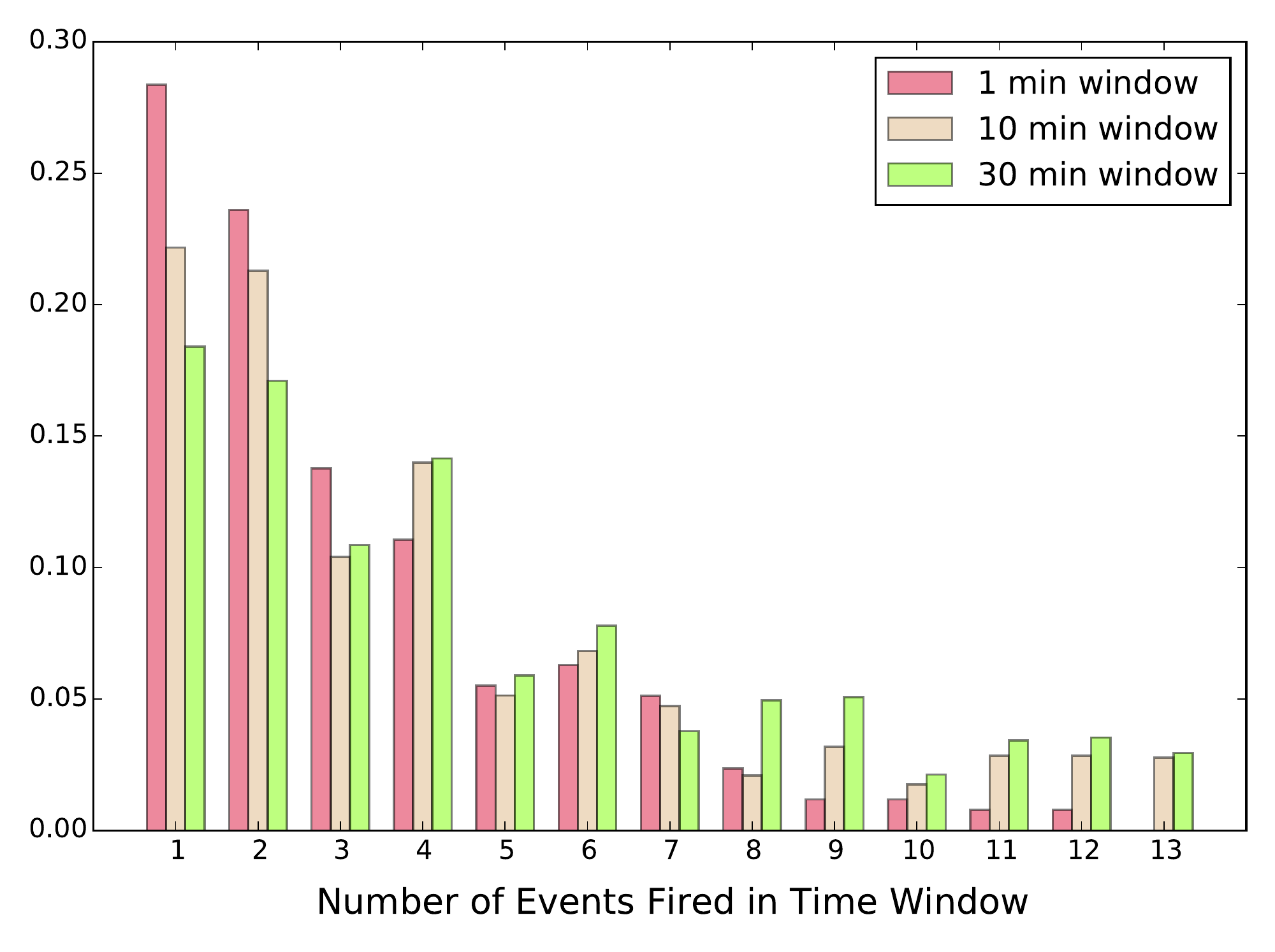}
  \caption{The distribution of the number of events that fire within a given time window. This is a long tail
    distribution as non-zero probabilities of event frequencies above 13 extend to over 100. These outliers are excluded
    from this histogram plot in order to highlight the high-probability frequencies. Time windows in which no events
    fire are not included in this plot.}
  \label{fig:firing-histogram}
\end{figure}

The events associated with each of the 4 modalities fire at very different rates as shown in
\tabref{firing-rate}. Moreover, text events fire in bursts, while the location events fire at regularly spaced intervals
when GPS signal is available. The app and web events fire at varying degrees of burstiness depending on the user.
\figref{firing-histogram} shows the distribution of the number of events that fire within each of the time windows. An
important takeaway from these distributions is that most events come in bursts followed by periods of inactivity. This
results in the counterintuitive fact that the 1 minute, 10 minute, and 30 minute windows have a similar distribution on
the number of events that fire within them. This is why the decrease in error rates attained from waiting longer for a
decision is not as significant as might be expected.

\begin{figure}[h]
  \centering
  \includegraphics[width=\columnwidth]{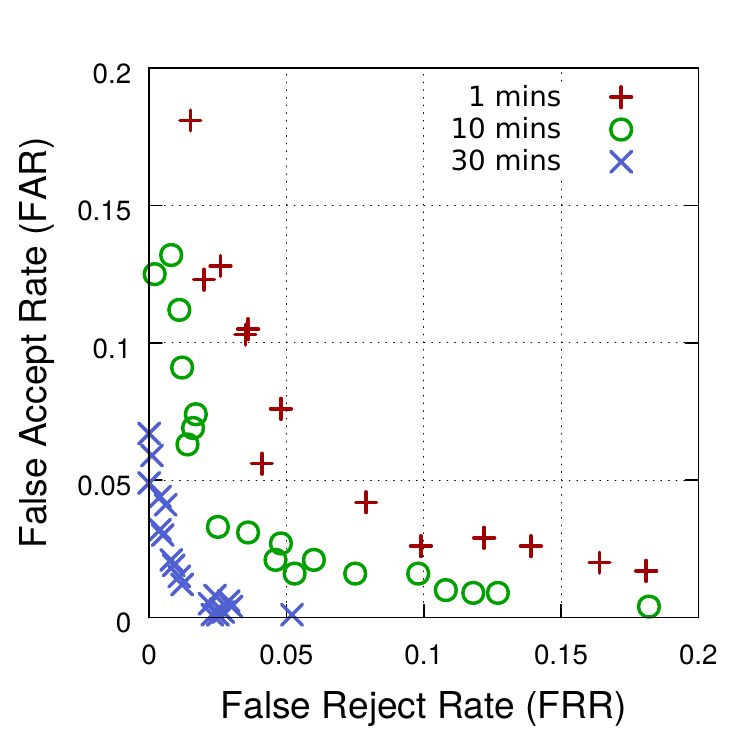}
  \caption{The performance of the fusion system with 4 classifiers on the 200 subject dataset. The ROC curve shows the
    tradeoff between FAR and FRR achieved by varying the threshold parameter $a_0$ in \eqref{fusion-rule}.}
  \label{fig:fusion-performance-pareto}
\end{figure}

Asynchronous fusion of classification of events from each of the four modalities is robust to the irregular rates at
which events fire. The decision fusion rule in \eqref{fusion-rule} utilizes all the available biometric data, weighing
each classifier according to its prior performance. \figref{fusion-performance-pareto} shows the receiver operating
characteristic (ROC) curve trading off between FAR and FRR by varying the threshold parameter $\tau$ in \eqref{Chair_Varshney_LRT}.

As the size of the decision window increases, the performance of the fusion system improves, dropping from an equal
error rate (EER) of 0.05 using the 1 minute window to below 0.01 EER using the 30 minute window.

\subsection{Contribution of Local Classifiers to Global Decision}\label{sec:contribution}

\begin{figure}[h]
  \centering
  \includegraphics[width=\columnwidth]{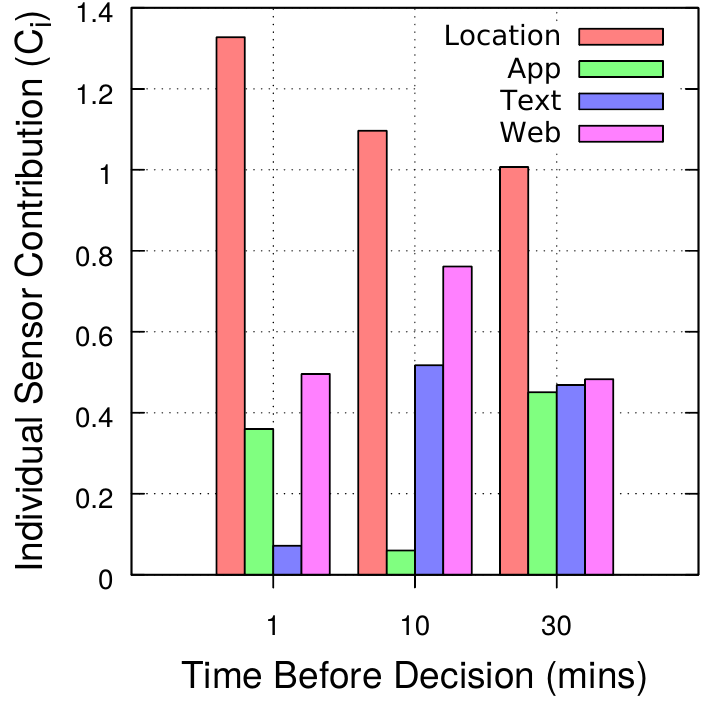}
  \caption{Relative contribution of each of the 4 classifiers computed according to \eqref{relative-contribution}.}
  \label{fig:individual-contributions}
\end{figure}

The performance of the fusion system that utilizes all four modalities of \textm{}, \appm{}, \webm{}, and \locationm{}
is described in the previous section. Besides this, we are able to use the fusion system to characterize the
contribution of each of the local classifiers to the global decision. This is the central question we consider in the
paper: what biometric modality is most helpful in verifying a person's identity under a constraint of a specific time
window before the verification decision must be made? We measure the contribution $C_i$ of each of the four
classifiers by evaluating the performance of the system with and without the classifier, and computing the contribution
by:

\begin{equation}\label{eq:relative-contribution}
  C_i=\frac{E_i - E}{E_i}
\end{equation}

\noindent where $E$ is the error rate computed by averaging FAR and FRR of the fusion system using the full portfolio of
4 classifiers, $E_i$ is the error rate of the fusion system using all but the $i$-th classifier, and $C_i$ is
the relative contribution of the $i$-th classifier as shown in \figref{individual-contributions}. We consider the
contribution of each classifier under three time windows of 1 minute, 10 minutes, and 30 minutes. Location contributes
the most in all three cases, with the second biggest contributor being web browsing. Text contributes the least for the
small window of 1 minute, but improve for the large windows. App usage is the least predictable contributor. One 
explanation for the \appm{} modality contributing significantly under the short decision window is that the first app
opened in a session is a strong and frequent indicator of identity. Therefore, its contribution is high for short
decision windows.

\section{Conclusion}\label{sec:conclusion}

In this work, we proposed a parallel binary decision-level fusion architecture for classifiers based on four biometric
modalities: text, application usage, web browsing, and location. Using this fusion method we addressed the problem of
active authentication and characterized its performance on a real-world dataset of 200 subjects, each using their
personal Android mobile device for a period of at least 30 days. The authentication system achieved an equal error rate
(ERR) of 0.05 (5\%) after 1 minute of user interaction with the device, and an EER of 0.01 (1\%) after 30 minutes. We
showed the performance of each individual classifier and its contribution to the fused global decision. The
location-based classifier, while having the lowest firing rate, contributes the most to the performance of the fusion
system.

\vspace{0.3in}

\noindent\textbf{Lex Fridman} is a Postdoctoral Associate at the Massachusetts Institute of Technology. He received his
BS, MS, and PhD from Drexel University. His research interests include machine learning, decision fusion, and numerical
optimization.

\noindent\textbf{Steven Weber} received the BS degree in 1996 from Marquette University in Milwaukee, Wisconsin, and the
MS and PhD degrees from the University of Texas at Austin in 1999 and 2003, respectively. He joined the Department of
Electrical and Computer Engineering at Drexel University in 2003, where he is currently an associate professor. His
research interests are centered around mathematical modeling of computer and communication networks, specifically
streaming multimedia and ad hoc networks. He is a senior member of the IEEE.

\noindent\textbf{Rachel Greenstadt} is an Associate Professor of Computer Science at Drexel University, where she
research the privacy and security properties of intelligent systems and the economics of electronic privacy and
information security. Her work is at "layer 8" of the network -- analyzing the content. She is a member of the DARPA
Computer Science Study Group and she runs the Privacy, Security, and Automation Laboratory (PSAL) which is a vibrant
group of ten researchers. The privacy research community has recognized her scholarship with the PET Award for
Outstanding Research in Privacy Enhancing Technologies, the NSF CAREER Award, and the Andreas Pfitzmann Best Student
Paper Award.

\noindent\textbf{Moshe Kam} received his BS in electrical engineering from Tel Aviv University in 1976 and MSc and PhD
from Drexel University in 1985 and 1987, respectively. He is the Dean of the Newark College of Engineering at the New
Jersey Institute of Technology, and had served earlier (2007-2014) as the Robert Quinn professor and Department Head of
Electrical and Computer Engineering at Drexel University. His professional interests are in system theory, detection and
estimation, information assurance, robotics, navigation, and engineering education. In 2011 he served as President and
CEO of IEEE, and at present he is member of the Boards of Directors of ABET and the United Engineering Foundation (UEF).
He is a fellow of the IEEE ``for contributions to the theory of decision fusion and distributed detection.''

\bibliographystyle{IEEEtran}
\bibliography{auth}

\end{document}